\documentstyle[aps,twocolumn,epsfig]{revtex}

\begin{document}

\date{\today}

\wideabs{\title{\bf Complex dislocation dynamics in ice: experiments}

\author{J\'er\^ome Weiss$^1$, Jean-Robert Grasso$^2$,
M.-Carmen Miguel$^{3}$,\\ Alessandro Vespignani$^{3}$, 
and Stefano Zapperi$^{4}$\\
$^1$LGGE CNRS, 54 rue Moliere
BP 96,\\38402 St Martin d'Heres Cedex , France\\
$^2$LGIT, Observatoire de Grenoble, BP 53x,\\
38041 Grenoble Cedex, France\\
$^3$The Abdus Salam International Centre for Theoretical
Physics\\
P.O. Box 586, 34100 Trieste, Italy\\
$^4$INFM, Universit\`a "La Sapienza", P.le A. Moro 2\\
        00185 Roma, Italy
}

\maketitle

\begin{abstract}
\noindent 
We present a statistical analysis of the acoustic emissions induced by 
dislocation motion during the creep of ice single crystals. The recorded
 acoustic waves provide an indirect measure of the inelastic energy 
dissipated during dislocation motion. Compression and torsion creep 
experiments indicate that viscoplastic deformation, even in the 
steady-state (secondary creep), is a complex and inhomogeneous 
process characterized by avalanches in the motion of dislocations. 
The distribution of avalanche sizes, identified with the acoustic 
wave amplitude (or the acoustic wave energy), is found to follow
 a power law with a cutoff at large amplitudes which depends
 on the creep stage (primary, secondary, tertiary). These results 
suggest that viscoplastic deformation in ice and possibly in other
 materials could be described in the framework of non-equilibrium
 critical phenomena.

\end{abstract}
}

\section{introduction}
ithin a material-dependent range of temperature and stress \cite{1}, 
the (visco)plastic deformation of crystalline materials involves 
the motion of a large number of dislocations. The importance of collective 
effects, through elastic interactions, on dislocation dynamics and the
 heterogeneous character of plasticity has been recognized for a long time 
(see e.g.\cite{2} ). Only relatively recently, 2D \cite{3,4}  as well as 
3D \cite{5}  
numerical 
simulations of collective dislocation interactions have revealed a spontaneous     self-organization     of dislocation patterns with walls and cells, 
in agreement with observations \cite{6} . However, collective dislocation
 dynamics has received much less attention, with the exception of 
the special case of Portevin-Le Chatelier (PLC) effect (see below). 
This may be due to the difficulty to monitor experimentally the variations
 of dislocation motion in time, energy and space domains. Here,
 we report an experimental analysis of acoustic emissions (AE)
 generated by dislocation motions which allow to reveal the 
intermittent and jerky character of collective dislocation dynamics.
 Experiments are performed on almost disorder free (except dislocations)
 ice single crystals. In a companion paper \cite{7}, we present numerical 
simulations which reproduce the main statistical characteristics of
 the AE measurements. This work (experiments and model) show that 
complex and jerky plastic flow is not restricted to the PLC effect, 
and rather the rule than the exception.

\section{Acoustic emission to monitor dislocation dynamics}
In solid materials, sudden local changes of inelastic strain 
generate AE waves \cite{8}. The sources can be crack nucleation and propagation,
 twinning, or dislocation motion. In the present work, AE is used to 
monitor dislocation dynamics. In our experiments, given the amplitude 
threshold and the frequency range of our transducer, the detected AE 
are unlikely to be the result of a single moving dislocation but most 
probably are related to synchronized accelerations of dislocations,
 called plastic instabilities or dislocation avalanches \cite{9}. 
From the theoretical analysis of Rouby et al. \cite{10} one can relate 
the maximum amplitude of the acoustic wave $A$ resulting from a 
plastic instability, to the number of involved dislocations $n$ and their
 velocity $v$ \cite{9}:  
\begin{equation}
A=k\frac{nLbvt_0}{d}
\end{equation}
where $k$ is a coefficient related to material properties and the 
piezoelectric constant of the transducer (dimension: V/m²), $b$ is the 
Burger' s vector, $L$ is the length of the n moving dislocations, 
$t_0$ is the travel time of the acoustic wave through the transducer 
(considered to be constant), and $d$ is the source/transducer distance 
(supposed to be large compared to $L$). In this crude model, $L$ and $v$ 
are supposed to be identical for all the involved moving dislocations. 
The dislocation velocity $v$ is considered to be zero before and after 
the event and constant during the event. The term $1/d$ represents 
the geometrical attenuation of the acoustic waves. 
From (1), Weiss and Grasso \cite{9} showed that the AE amplitude $A$ 
is a measure of the local strain associated with the dislocation avalanche. 
Therefore, the rate of global AE activity is an indirect measure of the 
global strain rate of the sample.

	The AE energy radiated by the acoustic wave is proportional 
to $A^2$. According to Kiesewitter and Schiller \cite{11}, 
the energy dissipated 
by viscoplastic deformation during an event also scales with $A^2$. 
This results from an expression given by Eshelby  \cite{12} for the energy 
dissipated at the source by a single screw dislocation of length $L$ 
moving at a velocity $v$:
\begin{equation}
E=KL^2b^2v^2
\end{equation}
where $K$ is a coefficient depending on material constants, 
including the shear modulus and the velocity of acoustic transverse waves, 
and $b$ the Burger's vector. A comparison of (1) and (2) with $n=1$ 
shows that $E \sim A^2$. This scaling has been observed during our
 experiments. 
Therefore AE allows to study during deformation the dislocation 
dynamics in energy, time, and possibly space (if AE sources locations 
are determined with the help of multiple transducers) domains.
In the present study, only two transducers were used, 
which did not allow 3D localization. 
\section{Experimental procedure}
As a model material to study dislocation dynamics from acoustic emission, 
ice provides the following advantages: single crystals \cite{13}  
or polycrystals 
with various microstructures can be easily grown in the laboratory; 
transparency allows verification that AE activity is not related to 
microcracking; and an excellent coupling between the ice and the AE 
transducers can be obtained by fusion/freezing. Within the range of 
temperature and stress corresponding to our experimental conditions, 
diffusional flow is not a significant mechanism of deformation in ice, 
and viscoplastic deformation occurs by dislocation motion \cite{14}. 
Hexagonal 
ice Ih presents a very strong plastic anisotropy of the single crystal
\cite{14}: 
viscoplastic deformation of single crystals occurs essentially by basal glide. 
	
Uniaxial compression creep as well as torsion creep experiments 
were performed at -$10~\circ$C on artificial ice single crystals 
(150 mm X 75mm), each of them constituted by several steps of constant 
applied stress. These experiments were similar to experiments performed 
previously and which are detailed in \cite{9} and \cite{15}. 
For the present new set of 
experiments, a new AE recording device was used which allowed to record, 
for each event, different characteristics including the arrival time 
(at a precision of 0.1 $\mu$s), the amplitude, the acoustic energy or the 
average frequency. Owing to a greater sensitivity of the recording device, 
much better statistics on AE events were obtained compared to previous 
work \cite{9,15}. 
The frequency band width of the transducers was 10-100 kHz. During an 
experiment the event amplitude threshold was adjusted to 30 dB, or 
$3×10^{-3}$ Volts, i.e. about 5 dB  above the noise level. The dynamic 
range between the amplitude threshold (3×$10~{-3}$ V) 
and the maximum recordable amplitude (10 V, or 100 dB) was 70 dB, 
i.e., 3.5 orders of magnitude. The corresponding dynamic range for
 energies was 7 to 8 orders of magnitude.

	Several independent evidences for recorded AE to be related
 to dislocation motion, including an absence of microcracks or a 
proportionality between the global strain-rate and the global 
AE rate have been detailed elsewhere \cite{9,15} .

\section{Statistical analysis of AE amplitudes and energies}

From equation (1), the acoustic activity $A(t)$ monitor the evolution of 
dislocation activity and viscoplastic dissipation. Figure 1 shows this 
evolution during the first loading step of a compressive creep test. 
It appears immediately that the instantaneaous viscoplastic dissipation 
is intermittent and jerky, showing bursts of activity that can be considered
 as avalanches of dislocations moving collectively in the material. 
Although the cumulated acoustic activity, i.e. an estimate of the strain, 
appears relatively smooth at the time scale of several thousands of seconds 
(Figure 1), At smaller time scales, or in terms of dissipated energy $E(t)$, 
this avalanche behavior is more obvious (Figure 2): dislocation dynamics
 during viscoplastic deformation is heterogeneous in time and energy domains. 
This recalls the jerky flow associated to Portevin-Le Chatelier effect 
occuring in different alloys (see e.g. \cite{16,17}).  
However, whereas PLC effect 
is attributed to the dynamic interaction of two defect populations, 
namely mobile dislocations and solute atoms \cite{17}, the avalanche behavior 
reported here results only from collective dislocation interactions 
(solid solutions in the artificial single crystals are below 10-10 mole 
frac.\ cite{13} ). For this reason, we believe that this kind of behavior should not
 be restricted to ice, as confirmed by a numerical modelling of this effect in a companion paper \cite{7}.
	
This avalanche behavior is recovered whatever the type of 
loading (compression or torsion), the applied stress, or the 
creep stage (primary, secondary, tertiary). However, interestingly and 
unexpectedly, the classical distinction between these creep stages 
corresponds also to different dislocation dynamics: primary and tertiary c
reep are associated to a pronounced jerky character and some very large events, whereas secondary creep appears smoother, at least at a large time scale.
 This difference, visible on figure 1, is clearer on the amplitude 
distributions. Secondary creep is characterized by power law statistics 
with an exponential cutoff at large amplitudes (or energies) (Figure 3):
\begin{equation}
N(A)\sim A^{-\tau}e^{-A/A_c}
\end{equation}
or the discrete distribution calculated with linear bins, or
\begin{equation}
N(A'>A)\sim A^{-\tau+1}e^{-A/A_c}
\end{equation}
for the cumulative distribution.
Because of the scaling $E \sim A^2$, similar relations are obtained for 
energy distributions with an exponent $(\tau+1)/2$. For secondary creep under 
compression, the exponent $\tau$ of 
the discrete distributions, estimated from a 
linear fit on a log-log plot below the amplitude cutoff $A_c$, is found to be
 stable through creep deformation, independent of the applied stress and 
equal to $2.0\pm 0.05$. Similarly, $A_c$ is found to be 
constant during a given 
loading step (Figure 3) and does not seem to vary significantly with the 
applied stress around $A_c \sim 0.4-0.5 V$. 
This  means that power law scaling is 
clearly observed over 1.5 orders of magnitude, i.e. 3 to 4 orders for  
energies. The stability of this scaling justifies, on the dynamical point 
of view, the term steady-state sometimes used to identify secondary 
creep.

On the other hand, statistics of dislocation avalanches change with the  
stage of creep. Primary as well as tertiary creep are characterized by a 
power law scaling without detectable upper cutoff over the experimental 
scale range ($A \simeq 10 V$). In other 
words, $A_c$, if any, is rejected towards much 
larger scales compared to secondary creep. This transition from 
secondary to tertiary creep is visible on figure 4. Slightly larger $\tau$
 exponents, around $2.15\pm0.05$ seem also to be associated with primary and 
tertiary creep. The variation of AC with the creep stage suggests that 
this upper cutoff is not a finite-size effect.
Under torsion creep, power law scaling was also observed without 
detectable upper cutoff, but with a smaller exponent, $\tau=1.7$. This
 difference of $\tau$ on the loading mode is not explained so far.

\section{Time patterning}
To study the time patterning of dislocation dynamics, we calculated 
the correlation integral $C(t)$:
\begin{equation}
C(t)=\frac{2}{N(N-1)}n(\Delta t<t)
\end{equation}
where N is the total number of events and $n(\Delta t<t)$ 
is the number of pairs 
of events (not necesseraly successive events) separated by less than a time
 t. $C(t)$ is therefore simply the probability for two acoustic events to 
be separated in time by less than t. Note that this analysis does not take 
into account the amplitude of the events. For a random Poisson process with 
uncorrelated (identical) events, $C(t)\sim t$. A typical profile of  $C(t)$ is given 
on figure 5. Whereas a scaling $C(t)\sim t$ 
is observed above a threshold $tc$, the 
events are clustered below tc (the probability $C(t)$ is larger than expected 
for uncorrelated events). This behavior was observed whatever the type of 
loading (compression/torsion), the applied stress or the creep stage, but 
$t_c$ was found to follow the evolution of $<\Delta t>$,
 the average time interval 
between two successive events. These observations, along with detailed 
examinations of the data files, can be explained as follows: dislocation 
avalanches consist of a mainshock correlated in time with 
a sequence of few aftershocks. Note that these shocks are
 distinct dynamical events characterized by distinct acoustic waves. At 
larger time scales, viscoplastic deformation consists of avalanches which 
are themselves uncorrelated in time. When the AE activity is very high, 
the frequency of avalanches is so high such that the aftershock sequences 
are not revealed by the correlation integral analysis.

\section{Discussion and conclusion}

	From a statistical analysis of acoustic emissions generated by 
dislocation motion, we have shown that:
(i) viscoplastic deformation during the creep of ice occurs by a 
succession of dislocation avalanches. This avalanche behavior, in 
very pure ice single crystals, results only from dislocation-dislocation 
elastic interactions.
(ii) dislocation avalanches consist of a mainshock correlated in time with 
a sequence of few aftershocks.  These avalanches are themselves uncorrelated 
in time.
(iii) the distributions of AE amplitudes (the ``size'' of the avalanche) 
or AE energies follow a power law scaling with an exponential upper cutoff 
which depends on the creep stage. For a specific creep stage, the statistical
 properties are stable and do not vary with time or the applied stress. 
Such power law scaling of dislocation avalanches suggests that 
viscoplastic deformation in ice and possibly in other materials 
could be described in the framework of non-equilibrium critical 
phenomena \cite{15}. 
In particular, driven-dissipative systems have the tendency to develop 
singular response functions and avalanche like behavior depending on the 
driving mechanism. This critical behavior is, in fact, resulting from the
 collective behavior of the many microscopic degrees of freedom of the 
system responding to the external perturbation. It is then of a major 
interest to develop theoretical models that can relate the observed
 experimental behavior with the microscopic dynamics of dislocations 
and its interactions with the external driving fields. In the companion
 paper [7] we provide an attempt in this direction by developing a 
numerical model that successfully reproduces the avalanche behavior
 of dislocation motion.
(iv) whatever the very nature of the involved critical phenomena, 
Weiss et al.  have shown that power law distributions of dislocation 
avalanches imply that large plastic instabilities account for most of 
the viscoplastic deformation, rather than independent movements of 
individual dislocations. 

	The results presented in a companion paper [7] represent 
a further evidence to argue that this avalanche behavior could not 
be specific to ice, rather a common feature of collective dislocation 
dynamics. 

\section*{Acknowledgments}
J. Weiss has been supported on this work by the Action Th\'ematique 
innovante of INSU. The AE recording device was bought with the help 
of the BQR program of Universit\'e J. Fourier, Grenoble I. Single 
crystals were prepared by O. Brissaud with an experimental setup 
designed by F. Domin\'e. We thank F. Lahaie for valuable discussions.

\begin{figure}
\centerline{\epsfig{file=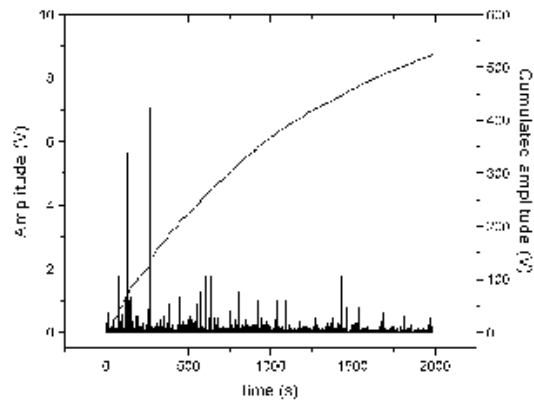,width=8truecm}} 
\caption{\small Instantaneous as well as cumulated acoustic activity 
(amplitude) during the first loading step of a compression creep 
test (applied stress=0.58 MPa).  Transition from primary to 
secondary creep.  }
\end{figure}

\begin{figure}
\centerline{\epsfig{file=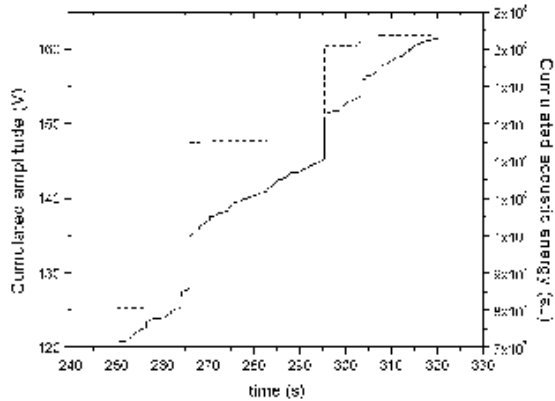,width=8truecm}} 
\caption{\small Detail of figure 1 within a time window: Cumulated amplitude 
(solid line) as well as cumulated energy (dashed line)  }
\end{figure}

\begin{figure}
\centerline{\epsfig{file=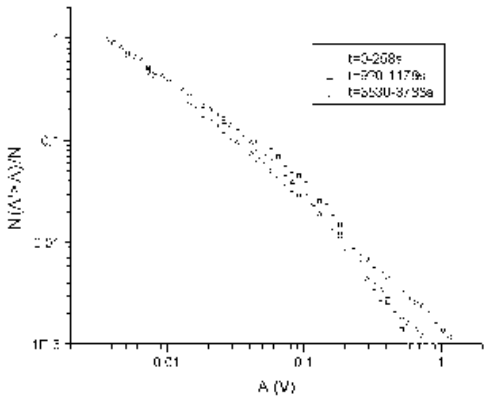,width=8truecm}} 
\caption{\small Cumulative amplitude distributions $N(A'>A)$, 
normalized by the total number of events in the time window, $N$, 
during secondary creep under compression (fourth step of loading; 
applied stress=1.29 MPa). The three sets of data (2500 events each, 
or a duration of about 110-120s) correspond to three non-overlapping 
time windows. The dashed line corresponds to $\tau=2$ 
($\tau-1=1$ for the cumulative distribution)}
\end{figure}

\begin{figure}
\centerline{\epsfig{file=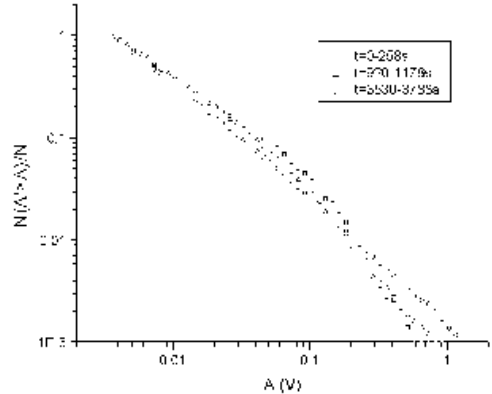,width=8truecm}} 
\caption{\small Cumulative amplitude distributions $N(A'>A)$, normalized by 
the total number of events in the time window, $N$, during the sixth step of 
loading of a compression creep test (applied stress=1.64 MPa): Transition 
from secondary to tertiary creep. The three sets of data correspond to 
three non-overlapping time windows of duration 258s. This duration 
corresponds to about 5000 events for two windows ( and )) 
and about 71000 events for the last one (F, tertiary creep).}
\end{figure}

\begin{figure}
\centerline{\epsfig{file=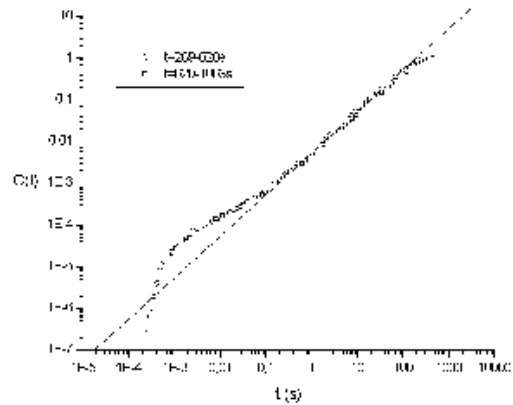,width=8truecm}} 
\caption{\small Correlation integral $C(t)$ calculated on two non-overlapping 
time windows (2500 events each) during the second loading step of a 
compression creep test  (applied stress=0.71 MPa). The dashed line 
corresponds to $C(t)\sim t$. The sharp decrease of $ C(t)$ below $5×10^{-3}$s 
results from the duration of the longuest events: the experimental 
device is unable to detect a new event as long as the former event 
is not finished. This limits artificially $C(t)$.}
\end{figure}
\end{document}